\documentclass[sigconf]{acmart}
\AtBeginDocument{%
  }

\setcopyright{none}                
\renewcommand\footnotetextcopyrightpermission[1]{} 
\makeatletter
\fancypagestyle{standardpagestyle}{%
  \fancyhf{}%
  \renewcommand{\headrulewidth}{0pt}%
  \fancyhead[R]{\@headfootfont\shortauthors}%
}
\makeatother

\usepackage{xargs}      
\usepackage{soul}       
\usepackage{color}      
\usepackage{xspace}     
\usepackage{xpunctuate} 
\usepackage{booktabs}
\usepackage{amsmath, amsfonts}
\usepackage{multirow}
\usepackage{algorithmic}
\usepackage{algorithm}
\usepackage{array}
\usepackage[caption=false,font=normalsize,labelfont=sf,textfont=sf]{subfig}
\usepackage{textcomp}
\usepackage{stfloats}
\usepackage{url}
\usepackage{verbatim}
\usepackage{graphicx}
\usepackage{wrapfig}
\usepackage{tcolorbox}
\tcbuselibrary{breakable}

\graphicspath{{figs/}{figures/}{pictures/}{images/}{./}} 

\newcommand{\ie}{{i.e.,}\xspace}
\newcommand{\eg}{{e.g.,}\xspace}

\newcommand{\bpstart}[1]{\vspace{0px}\noindent\textbf{#1}}

\definecolor{lightpink}{RGB}{237,157,202}
\definecolor{lightred}{RGB}{210,121,121}
\definecolor{lightorange}{RGB}{230,170,50}
\definecolor{lightgold}{RGB}{210,194,121}
\definecolor{lightgreen}{RGB}{121,210,121}
\definecolor{lightaqua}{RGB}{121,206,210}
\definecolor{lightblue}{RGB}{121,124,210}
\definecolor{lightpurple}{RGB}{153,102,255}
\definecolor{red}{RGB}{178,34,34}
\definecolor{gray}{RGB}{166,166,166}
\definecolor{bluebox}{RGB}{198, 230, 255}
\definecolor{redbox}{RGB}{255, 212, 205}

\newtcolorbox{infobox}{
    width=\linewidth,
    colback=bluebox, 
    boxsep=-1pt,
    left=10pt,
    boxrule=0.25pt,
    before skip=2mm,
    after skip=2mm,
    breakable
}

\newcommandx{\guest}[3][1=]
    {\setulcolor{lightorange}{\ul{#1}} \textcolor{lightorange} 
    {[\textbf{#2:} #3]}}
\newcommandx{\luke}[2][1=] 
    {\setulcolor{lightred}{\ul{#1}} \textcolor{lightred} 
    {[\textbf{Luke:} #2]}}  
\newcommandx{\jeff}[2][1=] 
    {\setulcolor{lightpurple}{\ul{#1}} \textcolor{lightpurple}  
    {[\textbf{Jeff:} #2]}}

\newcommand{\figureteaser}{
    \begin{figure*}
        \centering
        \includegraphics[width=\linewidth]{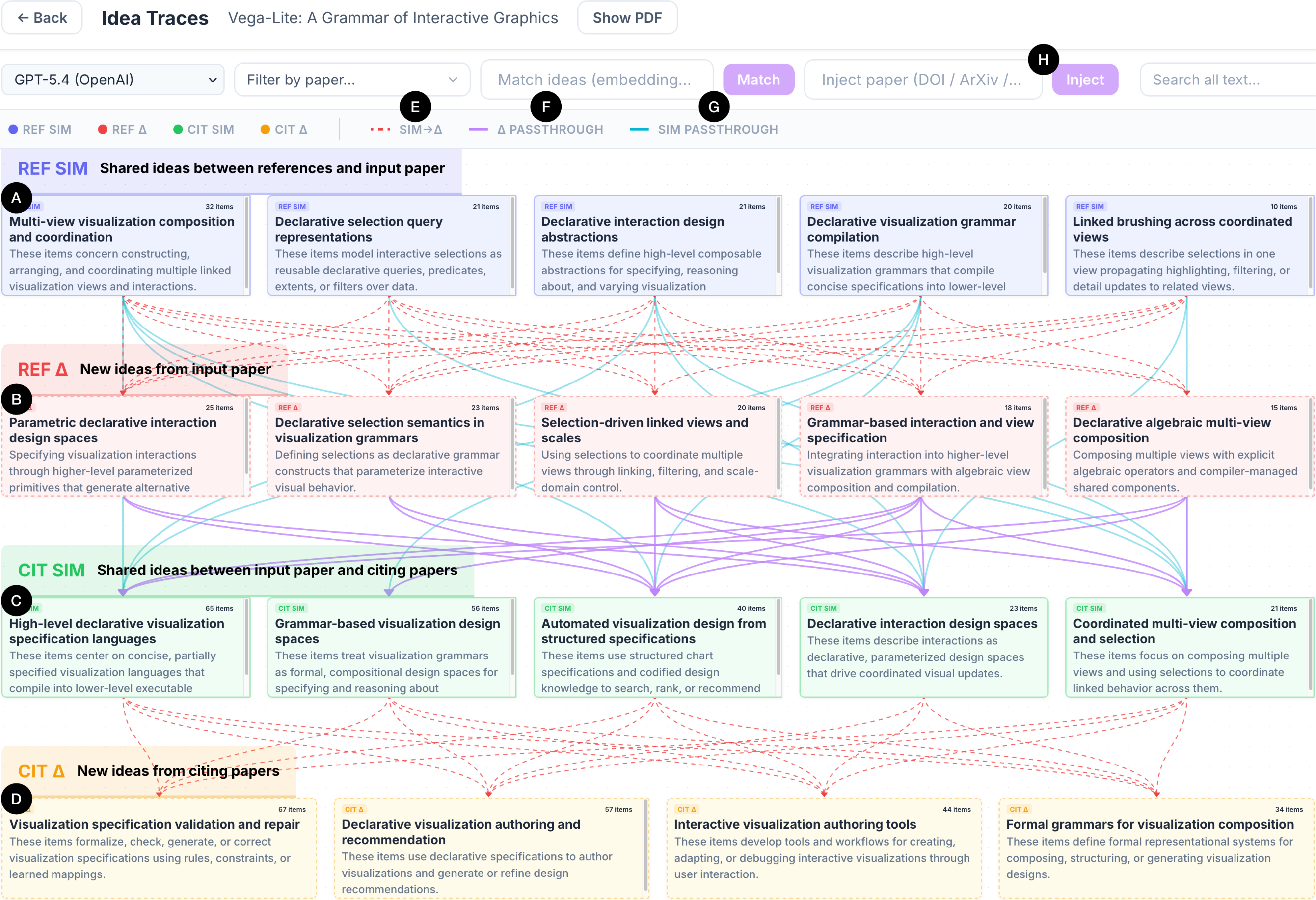}
        \vspace{-4mm}
        \caption{Interactive layout of the generated idea dependency graph for Vega-Lite~\cite{satyanarayan2016vega}.
        Idea traces enable fine-grained inspection of how ideas relate and develop across a network of papers: we can observe how Vega-Lite draws on more specialized declarative and interaction representations in the top-most layer to influence a wide range of downstream applications in the bottom-most layer, such as chart recommendation and authoring tools.
        Idea layers: clusters of (A) similarities (shared ideas) between Vega-Lite and its references, (B) deltas (new ideas or differences) from Vega-Lite, (C) similarities between Vega-Lite and subsequent citing papers, and (D) deltas from citing papers. Passthrough connections: (E) red dashed edges indicate branching behavior from a similarity to delta idea, (F) purple edges indicate transitional ideas (i.e., deltas from Vega-Lite used in citing papers), and (G) blue edges indicate idea bases (i.e., similarities between Vega-Lite and its references used in citing papers). (H) New papers can be injected into the graph to support comparison of ideas from different communities.}
        \label{fig:teaser}
        \vspace{-1mm}
    \end{figure*}
}

\newcommand{\figurepipeline}{
    \begin{figure*}
        \centering
        \includegraphics[width=\linewidth]{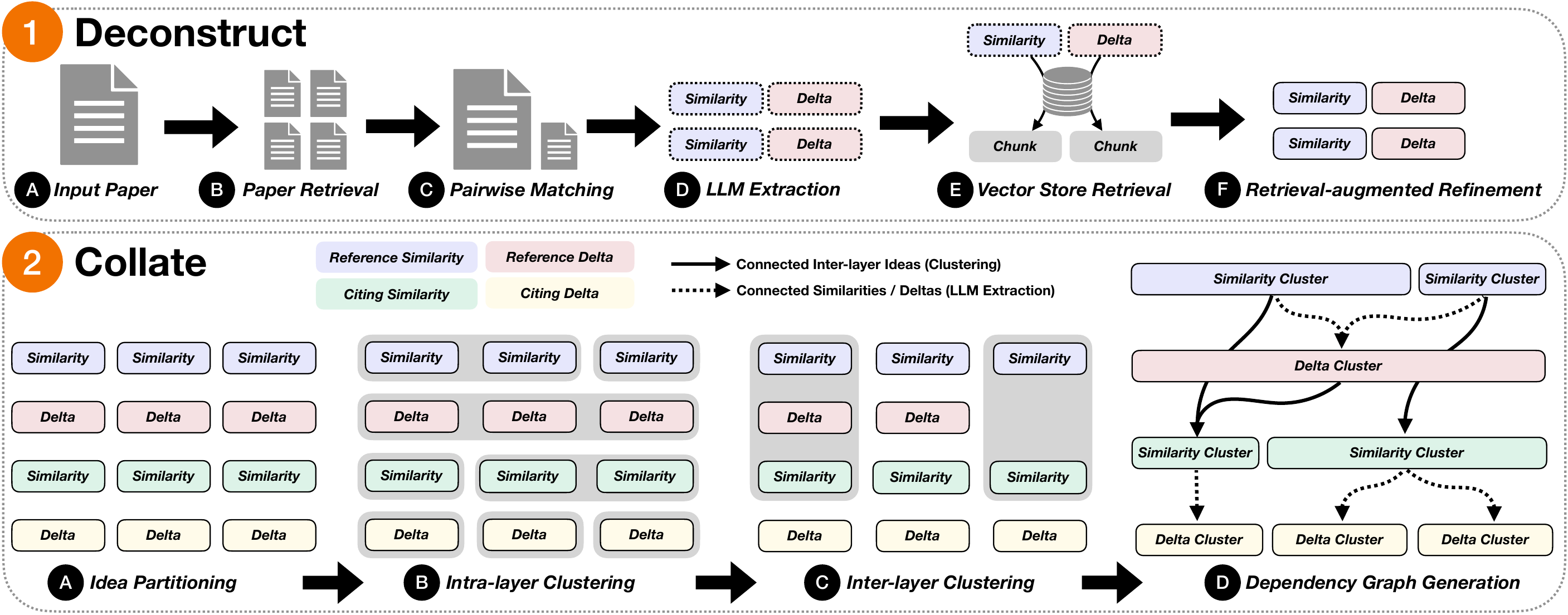}
        \vspace{-6.5mm}
        \caption{Overview of LLM pipeline. (1) Deconstruction: (A) a seed paper is provided as input, (B) the input paper's 1-hop references and citing papers are retrieved, (C) the input paper is paired with each retrieved paper, (D) the LLM extracts salient similarities and deltas from each pair, (E) the most relevant chunk for each extracted idea is retrieved from the vector store, and (F) the extracted ideas are paired with retrieved chunks for a final round of LLM refinement. (2) Collation: (A) Refined ideas are partitioned into four layers based on idea and citing membership, (B) ideas \textit{within} each layer are clustered to capture related ideas, (C) ideas \textit{between} layers are clustered to surface idea bases and transitional forms, and (D) the dependency graph is generated with intra-layer clusters (from part 2B) as nodes and inter-layer clusters (from part 2C) and similarity $\rightarrow$ delta branches (from part 1D) as independent edges.}
        \label{fig:pipeline}
        \vspace{-1mm}
    \end{figure*}
}

\newcommand{\figurevlcasestudy}{
    \begin{figure*}
        \centering
        \includegraphics[width=\linewidth]{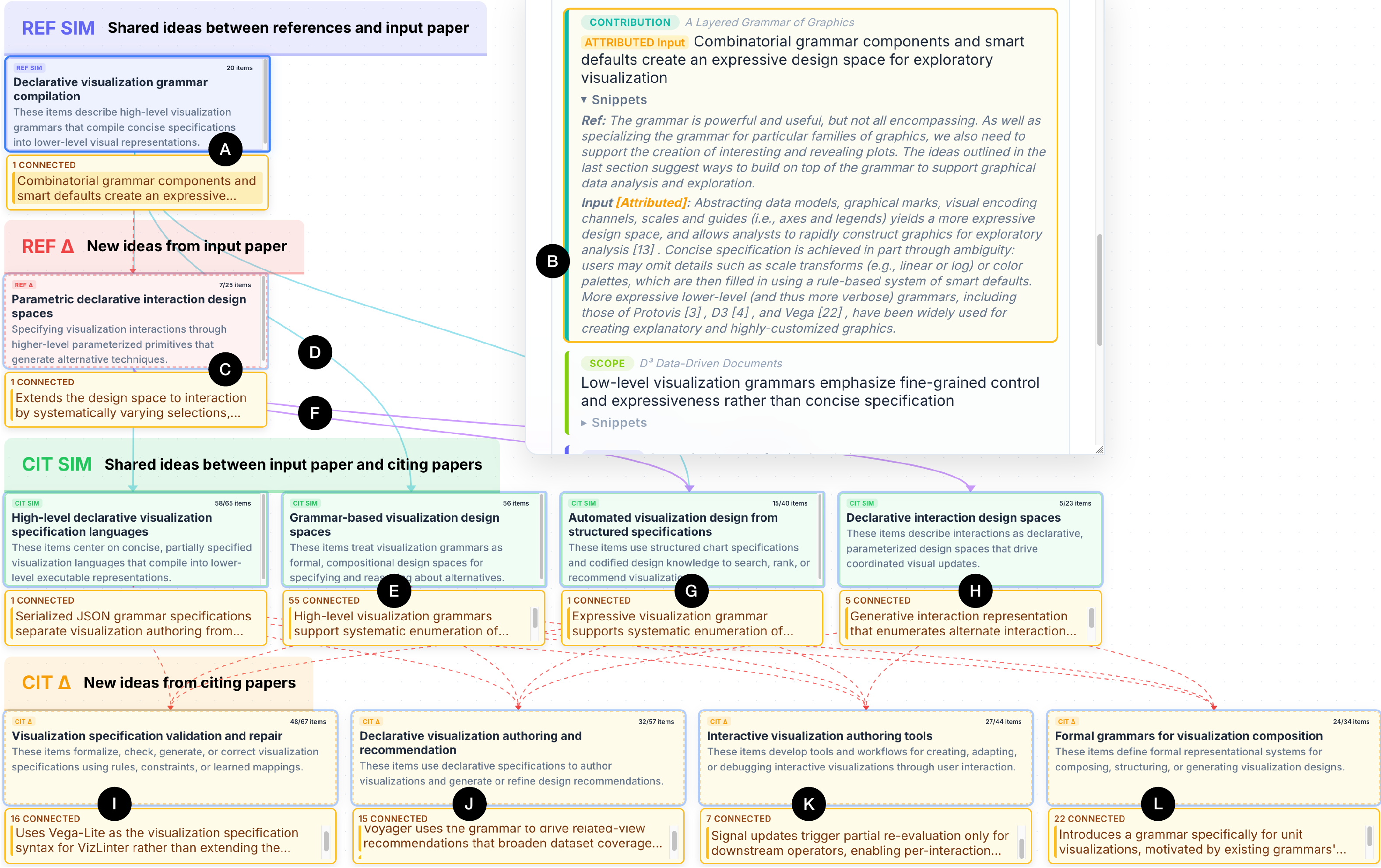}
        \caption{Overview of Vega-Lite's~\cite{satyanarayan2016vega} idea dependency graph after selecting an upstream idea: (A) the selected ``origin'' idea (a grammar-generated design space) from the top-most reference similarity layer, which propagates to several downstream ideas, (B) the detail view card, which surfaces text from both papers as supporting evidence (with the ``Attributed'' label highlighting Vega-Lite's explicit reference to prior work), (C) the delta from Vega-Lite that specializes this idea to interaction, (D) a passthrough edge from the origin idea to grammar design spaces in citing papers (E), indicating its influential role as an idea ``basis'', (F) passthrough delta edges, indicating ``transitional'' connections to citing hull similarities that leverage the new delta branch directly (G, H), and (I-L) new ideas in citing papers that can be traced directly back to the selected idea (A).}
        \label{fig:case-study}
    \end{figure*}
}

\newcommand{\figurenotebooklm}{
    \begin{figure}
        \centering
        \includegraphics[width=\linewidth]{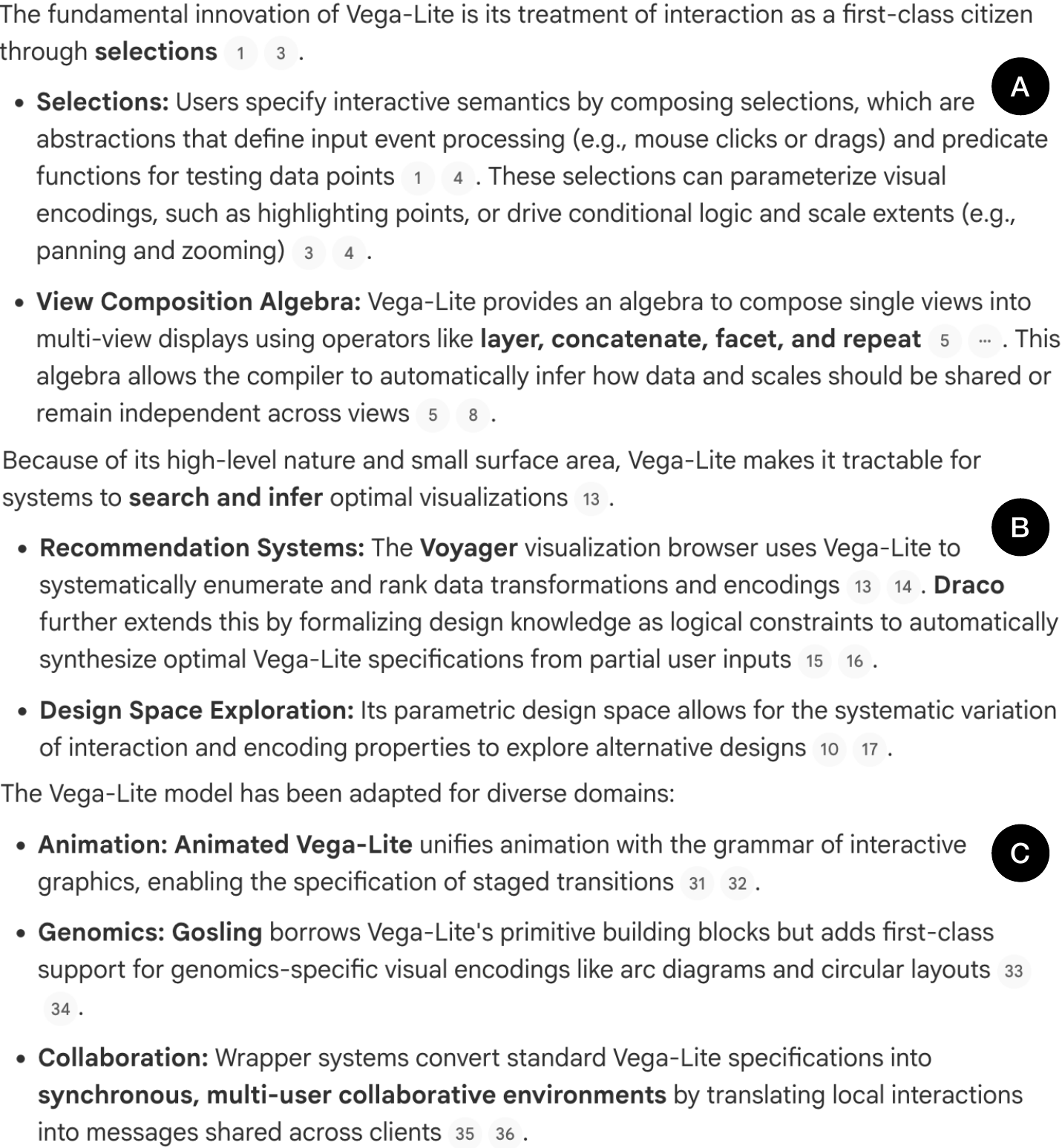}
        \caption{Summary of Vega-Lite's~\cite{satyanarayan2016vega} ideas from NotebookLM~\cite{notebooklm}: (A) core idea of a \textit{selection}, (B) new affordances for systematic search and inference over a design space, and (C) subsequent extensions to domains like animation. }
        \label{fig:notebook-lm}
        \vspace{0mm}
    \end{figure}
}

\newcommand{\figureinjection}{
    \begin{figure*}
        \centering
        \includegraphics[width=\linewidth]{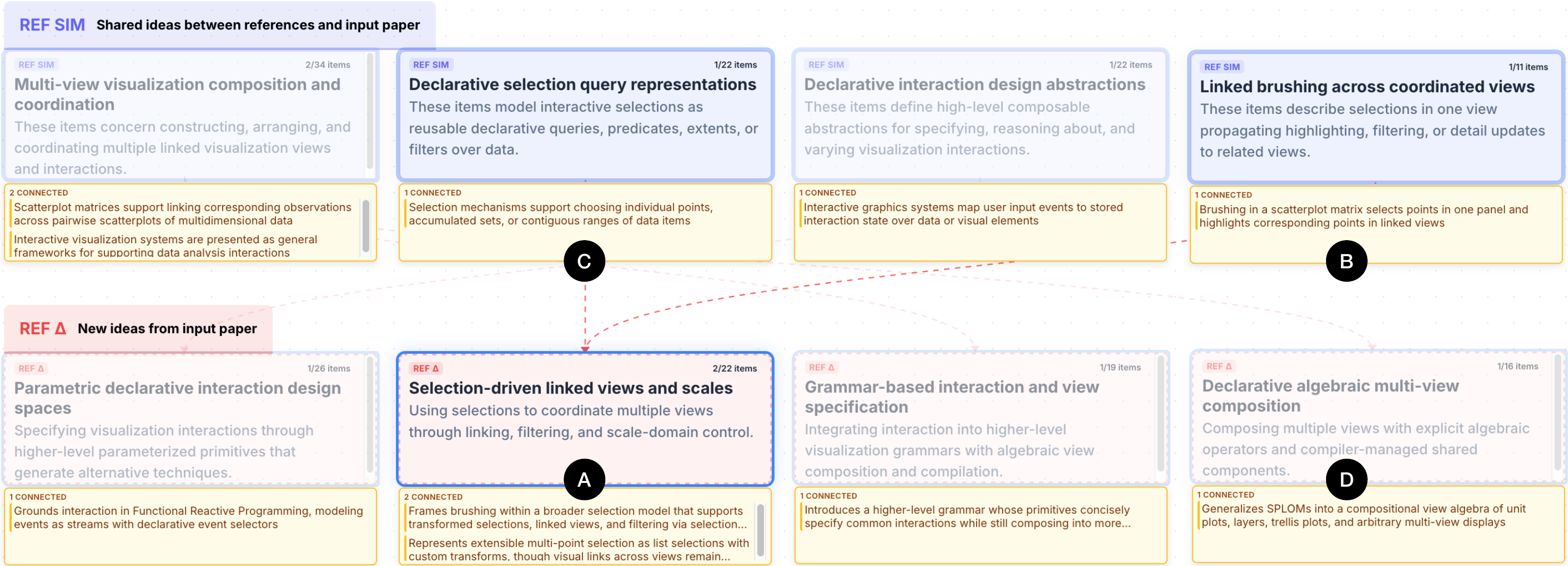}
        \caption{Injecting ``Brushing scatterplots''~\cite{becker1987brushing}, an influential paper not present in the reference hull, into Vega-Lite's idea graph: Vega-Lite's \textit{selection} delta (A) integrates two main prior ideas, recasting brushing and linking (B) as a selection mechanism (C). This abstraction allows Vega-Lite to generalize SPLOMs (scatterplot matrices) to other compositional views (D). }
        \label{fig:injection}
        \vspace{2mm}
    \end{figure*}
}

\newcommand{\figurecapturedistribution}{
    \begin{figure}[!t]
        \centering
        \includegraphics[width=0.8\linewidth]{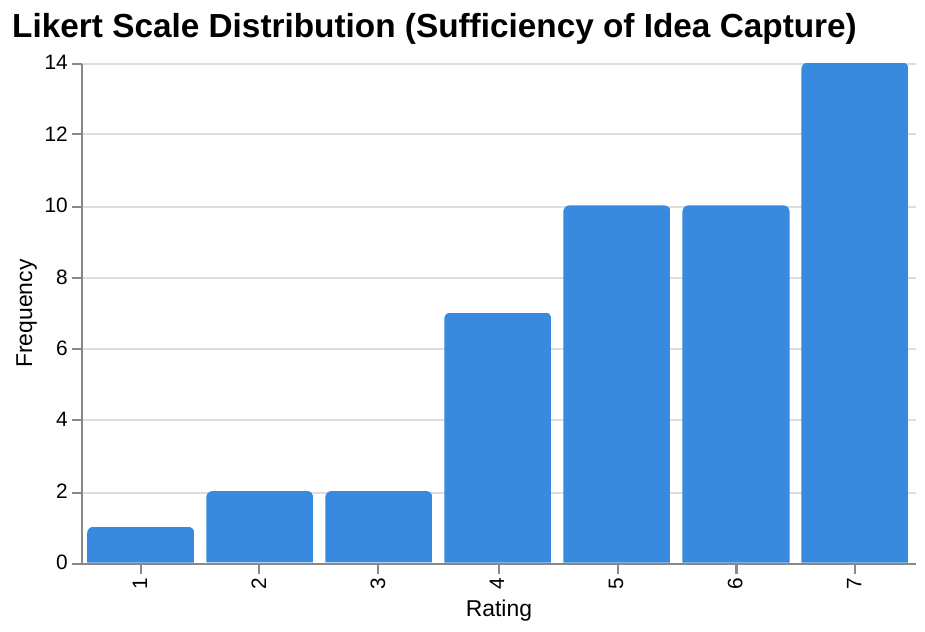}
        \caption{Task 2 (Idea Capture): Distribution of Likert scale responses.}
        \label{fig:capture-distribution}
    \end{figure}
}

\newcommand{\figuresimilaritydistribution}{
    \begin{figure}[!t]
        \centering
        \includegraphics[width=\linewidth]{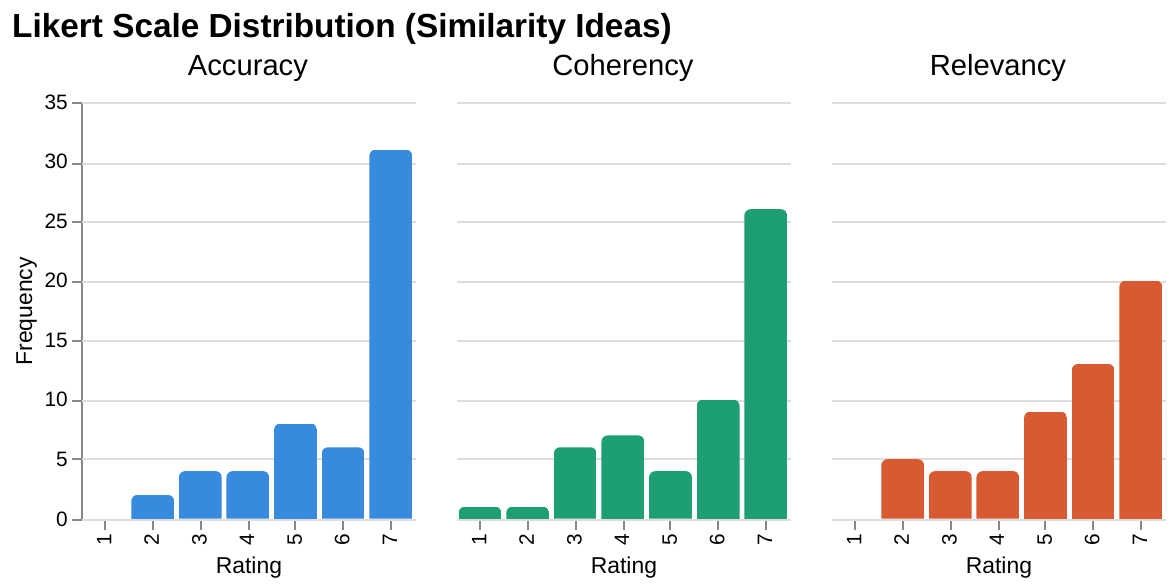}
        \caption{Task 3 (Idea Quality): Distribution of Likert scale responses for similarities.}
        \label{fig:similarity-distribution}
    \end{figure}
}

\newcommand{\figuredeltadistribution}{
    \begin{figure}[!t]
        \centering
        \includegraphics[width=\linewidth]{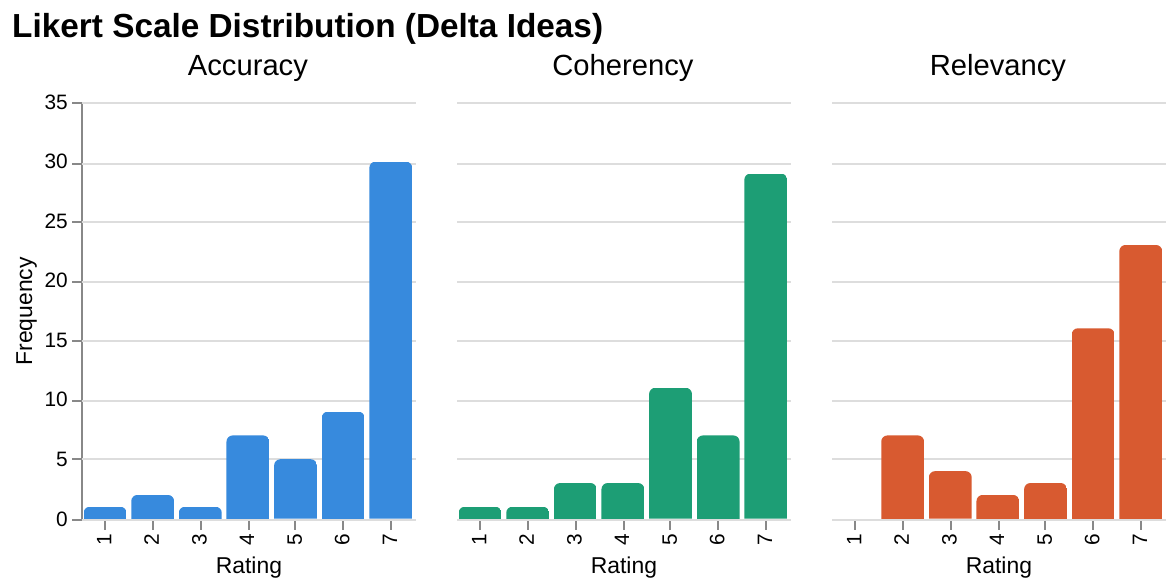}
        \caption{Task 3 (Idea Quality): Distribution of Likert scale responses for deltas.}
        \label{fig:delta-distribution}
    \end{figure}
}

\newcommand{\figuresurveydistribution}{
    \begin{figure*}[!t]
        \centering
        \includegraphics[width=\linewidth]{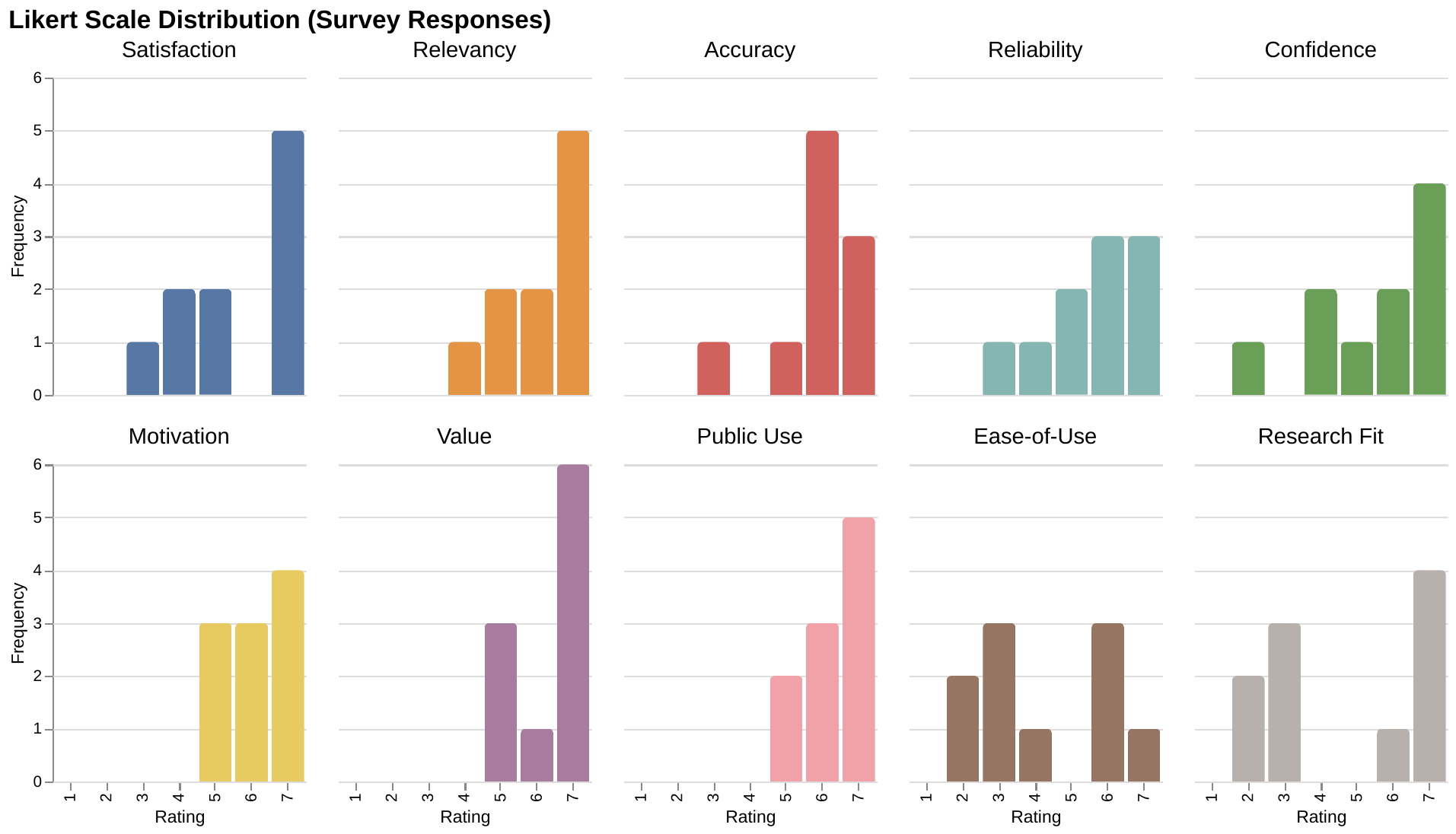}
        \caption{Post-interview Survey: Distribution of Likert scale responses.}
        \label{fig:survey-distribution}
    \end{figure*}
}

\newcommand{\tableparticipants}{
    \begin{table*}
    \centering
    \setlength\tabcolsep{0pt} 
    \smallskip 
    \small
    
    \def\arraystretch{1.2}%
    \scalebox{1}{
    \begin{tabular*}{\linewidth}{@{\extracolsep{\fill}} lllp{5in}l}
        \toprule
         & Age & Gender & Research Background & Selected Paper \\
        
        \midrule 
        P1 & 24 & Non-binary & 3rd year in Information Science PhD focusing on HCI, 6 years research experience & \cite{chopra2022negotiating}\\
        
        P2 & 38 & F & Final year PhD student in HCDE (Human Computer Design \& Engineering). 8 years experience, mostly qualitative and design research but with some quantitative and technical knowledge. & \cite{lee2005between} \\
        
        P3 & 28 & M & Machine learning, robotics, Neuro-inspired AI, 5th year PhD & \cite{wu2025mmsearch} \\
        
        P4 & 27 & F & PhD in Computer Science in HCI / AI, 5th year, 2 years of research experience before PhD & \cite{kazemitabaar2024improving} \\
        
        P5 & 28 & M & PhD, Computer Systems/Architecture, 5th year, 7 years research experience & \cite{hower2014heterogeneous} \\
        
        P6 & 29 & M & 6th year phd student in databases. Research focuses on video data management. 7 years of research experience. & \cite{bastani2020miris} \\

        P7 & 29 & F & 6th yr phd student in HCI, social computing. Also have 2-3 yrs of research experience before that. & \cite{wang2024investigating} \\

        P8 & 32 & F & 2nd year PhD student, Data visualization \& NLP, engineer in industry research before that for 7 years & \cite{reif2019visualizing} \\

        P9 & 27 & M & CSE, HCI, 6 years in PhD, 6 years total research experience & \cite{wang2024farsight} \\

        P10 & 26 & M & 5th year of PhD, ML, 6 year program with total research experience of 6 years. & \cite{gadre2023datacomp} \\
        
        \bottomrule
    \end{tabular*}
    }
    \vspace{1mm}
    \caption{ 
        Anonymized Participant Demographics \& Background.
    }
    \label{table:participants}
    \end{table*}
}

\begin{document}

\title{Idea Traces: Threading Accretive Content across Scientific Papers}


\author{Luke S. Snyder}
\email{snyderl@cs.washington.edu}
\affiliation{%
  \institution{University of Washington}
  \city{Seattle}
  \state{Washington}
  \country{USA}
}

\author{Jeffrey Heer}
\email{jheer@cs.washington.edu}
\affiliation{%
  \institution{University of Washington}
  \city{Seattle}
  \state{Washington}
  \country{USA}
}
%







\renewcommand{\shortauthors}{Luke S. Snyder and Jeffrey Heer}
\begin{abstract}
We introduce \textit{idea traces}, a novel thread-based representation of accretive content across scientific papers.
Rather than capturing fixed themes from papers, idea traces surface \textit{similarities} and \textit{deltas} of faceted content that enable inspection of how ideas relate and evolve across the citation network.
To extract traces, we contribute an LLM pipeline that deconstructs salient transitions between pairs of papers and collates the output into an idea dependency graph.
We leverage idea traces in a visual system that supports idea-driven navigation starting from a seed paper, lifting the ``plane'' of synthesis from papers to ideas.
To evaluate the usefulness and quality of generated ideas, we provide a case study that compares our approach with an LLM Q\&A workflow, and interview 10 researchers who used our system and annotated 110 ideas.
Experts found that our approach helped extract a useful space of connections that would have been difficult to make otherwise. 
Drawing on their feedback, we outline several promising directions for future research.
\end{abstract}

\maketitle
\makeatletter
\fancypagestyle{standardpagestyle}{%
  \fancyhf{}%
  \renewcommand{\headrulewidth}{0pt}%
  \fancyhead[LO]{\@headfootfont\shorttitle}%
  \fancyhead[RE]{\@headfootfont\shortauthors}%
}
\makeatother
\pagestyle{standardpagestyle}

\section{Introduction}
Scholarly synthesis remains a difficult and time-consuming task. 
Increasing specialization~\cite{head2021augmenting, hillesund2010digital, schema1985physicists, palani2023relatedly} and research output~\cite{bornmann2015growth, jinha2010article} require cognitively demanding workflows to locate and triage relevant prior work, comprehend key ideas and their relationships, and integrate findings from new papers.
Though prior work has addressed each stage of this process -- from dedicated interfaces for paper reading~\cite{head2021augmenting, gu2025abstractexplorer, badam2018elastic, masson2023charagraph, fok2023scim} to large-scale citation management systems and visual overviews~\cite{chau2011apolo, heimerl2015citerivers, shahaf2012metro, fang2025exploring} -- papers remain isolated ``silos'' of information.
Specifically, these techniques overlook the \textit{accretive} nature of scientific knowledge, leaving readers to synthesize relationships across the literature themselves~\cite{pirolli2005sensemaking}.

\figureteaser

We argue for a core reworking of how researchers read and explore papers that shifts from traditional paper-~\cite{head2021augmenting, fok2023scim, fok2024qlarify} and citation-based workflows~\cite{rachatasumrit2022citeread, kang2023synergi, basti2025citepeek, chang2023citesee} to idea-level search and analysis, lifting the ``plane'' of scholarly synthesis from papers to ideas.
This shift integrates two important avenues of prior work: thread-based literature synthesis~\cite{kang2022threddy, kang2023synergi} and research ideation support~\cite{pu2025ideasynth, radensky2024scideator, shahid2025literature, garikaparthi2025iris, li2025review}.
Existing thread-based workflows, which operationalize threads as specific lines of research (e.g., a section of related work), are useful for curating and organizing large collections of papers, 
though citation contexts are typically the unit of retrieval and analysis~\cite{kang2023synergi, rachatasumrit2022citeread, basti2025citepeek, chang2023citesee}.
Citances may fail to capture more nuanced, conceptual relationships outside their context, such as new framings of prior work, shared lineages of motivating problems, or divergent applications of an existing method.

Tools for ideation support~\cite{pu2025ideasynth, radensky2024scideator, shahid2025literature} have demonstrated that facet-based representations of a paper's ideas, such as the method, problem/motivation, and evaluation protocol, can serve as useful abstractions for inputs to Large Language Models (LLMs) to help ground and search ideas across papers.
Further, LLM chatbots and dedicated reasoning modules in AI-backed systems (e.g., Google NotebookLM~\cite{notebooklm}, Perplexity~\cite{perplexity}, Semantic Scholar~\cite{kinney2023semantic}, etc.) can synthesize key ideas within individual papers and across fields of research, serving as a useful entry point for scholars at the start of their literature review process.
However, cross-paper synthesis often remains a manual process since fine-grained ideas and their relationships are not provided as navigable outputs.

We contribute \textit{idea traces}, a novel thread-based representation of accretive content across scientific papers.
Idea traces integrate and extend prior thread- and idea-based literature tools, surfacing fine-grained \textit{threads} of \textit{faceted} content as they evolve across a network of papers (Fig.~\ref{fig:teaser}).
Specifically, an idea externalizes precise dimensions (e.g., problem, method, finding, etc.) that transmit between papers in the form of \textit{similarities} (i.e., a shared facet) and \textit{deltas} (i.e., a new facet that develops or extends a similarity in some meaningful way).
This representation both highlights salient features of a paper, by contextualizing it relative to prior/future work, and provides human-interpretable scaffolding for identifying shared intellectual bases and critical points of divergence.
Notably, ``trace'' signifies both meanings of the term: to track an individual idea's trajectory from paper to paper, and as a surviving artifact, impressed and molded from earlier forms.

We provide an LLM pipeline (Fig.~\ref{fig:pipeline}) to extract and collate idea traces, capturing aggregated relationships of shared ideas and their propagation.
Our pipeline deconstructs salient ideas via pairwise analysis of papers, using retrieval-augmented generation to refine and ground extracted output.
We perform a series of clustering operations to merge related ideas, using the captured similarities and deltas to partition into edge-directed layers across the citation hull. 
Clusters of ideas \textit{between} layers further materialize propagation for tracing: important differences (deltas) from a paper can be directly mapped to ideas that future citing work draws on (similarities). 
To support idea-based navigation and synthesis, we contribute an interactive interface that visualizes the resulting dependency graph as a node-link layout.
Scholars can input a paper of interest, select and navigate connections across the graph to trace the origins and different branches of an idea, and inject papers into the graph to surface potential relationships outside the citation network.

We evaluate our system through a case study and semi-structured interviews with 10 expert researchers.
Our case study examines a baseline LLM Q\&A system (Google NotebookLM~\cite{notebooklm}), noting differentiating aspects from our approach that afford new forms of representation, navigation, and analysis.
To examine the quality of generated ideas, experts annotated the coverage of key ideas from an input paper that they had read closely before, rated a total of 110 sampled similarities and deltas to assess quality and gradations of reasoning ability and failures, and explored a paper they were unfamiliar with using our system.
Experts found coverage to be reasonable for a sizable portion of their ideas (74\% with \text{$\geq$5} ratings on a 7-point Likert scale), with a majority (52\%) still receiving sufficient coverage (\text{$\geq$6}).
The quality of generated ideas was also favorable.
Across all ratings, partial alignment scores (\text{$\geq$5}) indicated useful information (81\% accuracy, 79\% coherency, and 76\% relevancy), with high alignment (\text{$\geq$6}) reported in the majority of instances (69\% accuracy, 65\% coherency, 65\% relevancy).

Experts frequently reported finding connections that were useful, interesting, and unexpected, helping triage specific content that would have been prohibitively difficult to locate otherwise.
Building off this feedback, we discuss important directions of future work, such as interactively refining input ideas to anchor on concepts of interest, extending search and extraction to author and community networks beyond citations, and supporting synthesis for downstream tasks (e.g., collaboration, paper writing, and personalized memory aids).
In summary, our primary contributions are:
\begin{enumerate}
    \item Idea traces, a novel thread-based representation of accretive content that models key relations as similarities and deltas.
    \item An LLM pipeline that extracts and collates a dependency graph to capture aggregated relationships of shared ideas.
    \item An interactive system for visualizing and tracing ideas across a collection of papers.
    \item A case study and interviews with 10 expert researchers, demonstrating that our approach provides a useful level of abstraction for cross-paper idea synthesis.
\end{enumerate}

\section{Related Work}
We draw from prior work on reading support and comprehension, as well as workflows that support cross-paper discovery and synthesis. 
We then situate our work relative to more recent AI-driven systems for idea generation and text analysis.

\subsection{Reading Augmentations}
Given the central role of reading to knowledge discovery and sensemaking, significant efforts have sought to enhance comprehension through augmented interfaces.
Relevant methods include intelligent tooltip features~\cite{head2021augmenting}, expandable abstracts~\cite{fok2024qlarify}, highlights that surface salient content~\cite{gu2024ai, lee2016spotlights, palani2023relatedly, fok2023scim, chi2005scenthighlights}, quizzes~\cite{richards2023readerquizzer}, or graphical overlays that present additional data as context~\cite{badam2018elastic, masson2023charagraph}.
A related line of research leverages citations to provide additional information, such as how later papers cite a focused paper~\cite{rachatasumrit2022citeread}, or use the reader's interaction history (e.g., bookmarked citations or paper alerts) to personalize results~\cite{chang2023citesee, lee2024paperweaver}.

These techniques can be characterized by the scope of information surfaced to the reader, from relevant text within the paper~\cite{fok2024qlarify, fok2023scim, richards2023readerquizzer, masson2023charagraph} to broader context across citation networks~\cite{chang2023citesee, rachatasumrit2022citeread, basti2025citepeek}.
We extend this notion to \textit{collections} of papers, extracting and collating salient information scattered throughout that would be difficult and time-consuming to discover otherwise.

\subsection{Scholarly Synthesis Workflows}
Synergi~\cite{kang2023synergi} frames workflows for scholarly synthesis as ranging from bottom-up, manual approaches~\cite{kang2022threddy, han2022passages, chau2011apolo} to top-down visual systems for high-level overviews~\cite{shahaf2012metro, heimerl2015citerivers, liu2024selenite, liu2025city, ma2025garden, ponsard2016paperquest}.
Manual techniques may leverage clipping and highlighting~\cite{han2022passages, kang2022threddy, chang2020mesh, rachatasumrit2021forsense, kuznetsov2022fuse} or hierarchical organization~\cite{kang2022threddy} to construct artifacts that serve as synthesis aids while reading multiple papers.
Top-down methods by comparison, such as automated LLM Q\&A systems~\cite{perplexity, notebooklm, taylor2022galactica}, visual overviews~\cite{shahaf2012metro, heimerl2015citerivers, liu2025city, ma2025garden}, or citation networks~\cite{chau2011apolo, heimerl2015citerivers, kang2022you}, can provide useful starting points for understanding key ideas or papers within a larger body of work.
Synergi~\cite{kang2023synergi} contributes a mixed-initiative workflow that straddles this space, providing AI-generated threads that can be inspected and modified to fit the user's needs.
Other techniques may exploit author networks~\cite{kang2022you, kang2023comlittee}, analogical reasoning and discovery~\cite{chan2018solvent, hope2017accelerating, kang2022augmenting, kang2025biospark}, or content overlap~\cite{palani2023relatedly, gu2025abstractexplorer} to highlight potentially relevant information.

Idea traces differ from these workflows in two primary aspects.
First, we move beyond citation contexts as the primary unit of analysis.
Prior systems often use citances as anchors to retrieve and organize content~\cite{kang2023synergi, chang2023citesee, basti2025citepeek, rachatasumrit2022citeread} (e.g., building a thread with similar context to a user-selected snippet~\cite{kang2023synergi}), though they lack additional nuance regarding the citation intent or broader set of potentially connected ideas.
For instance, Synergi's~\cite{kang2023synergi} citances of Threddy~\cite{kang2022threddy} do not discuss the more central connection of \textit{threads} as the organizational unit for synthesis, in contrast to approaches before them.
Further, Synergi's discussion of boundary objects does not cite the seminal paper from sociology~\cite{star1989institutional}, suggesting that this idea has become part of the HCI vernacular.
Workflows reliant on citation contexts alone may, therefore, fail to capture subtle or even ``obvious'' connections between papers that provide useful scaffolding to relate key ideas.

Second, we surface the \textit{fine-grained} development of ideas, moving beyond AI synthesis workflows~\cite{liu2024selenite, fok2025facets, newman2024arxivdigestables, chen2023marvista} and chatbots.
For instance, tools such as Google NotebookLM~\cite{notebooklm} and Perplexity~\cite{perplexity} can synthesize key, high-level ideas from a collection of papers. 
However, without access to their traces, readers lack concrete context for understanding the history of an idea, its variations, and connection to future work.
This limits readers' ability to identify and track niche ideas, assess an idea's novelty or influence, or generate new ideas from a large set of papers, all of which are supported by idea traces.

\subsection{LLM Systems for Ideation \& Text Analysis}
Several new lines of work have explored how to assist scholars in generating novel ideas, leveraging LLMs to facilitate exploration and creativity.
For instance, IdeaSynth~\cite{pu2025ideasynth} provides facet-based exploration of ideas in a graphical interface, visualized as a node-link diagram.
Users can iteratively expand facets to suggest new variations (e.g., a different method or problem area) and combine them to surface potentially novel instances.
A related system, Scideator~\cite{radensky2024scideator}, also leverages facet representations to ground the novelty of an input idea.
Here, novelty is defined by at least one variation of a facet, which can be used to query the Semantic Scholar API~\cite{kinney2023semantic} to locate overlapping papers.
We take inspiration from this line of work, leveraging fine-grained facets as ``building blocks'' to scaffold LLM deconstruction and human interpretation, since extracted facets can be efficiently verified and compared with the source text.

We also draw from an emergent family of LLM-backed data processing systems that target unstructured text~\cite{liu2025palimpzest, shankar2024docetl, shankar2025steering, lam2024concept}.
Central to this thread of research is a core set of abstractions for modeling intermediate representations that help generate structured outputs.
Mirroring the concept induction algorithm from LLooM~\cite{lam2024concept}, we provide an analogous set of operators that filter salient content (Fig.~\ref{fig:pipeline}: Steps 1C,D), refine ambiguous outputs (Fig.~\ref{fig:pipeline}: Step 1F), and cluster items to capture meaningful abstractions (Fig.~\ref{fig:pipeline}: Steps 2B,C).


\figurepipeline

\section{Idea Traces}
Idea traces externalize \textit{similarities} and \textit{deltas} of faceted content.
To extract traces, we instrument an LLM pipeline (Fig.~\ref{fig:pipeline}) to localize the transmission of similarities and deltas between pairs of papers, and collate and synthesize the output into an idea dependency graph.
This graph enables readers to identify similar ideas scattered across papers (that may lack explicit citing relationships) and trace their development -- \ie how subsequent papers extend or change an idea (the delta).
In the following sections, we discuss our motivations and rationale that led us to select these representations and then describe the pipeline.

\subsection{Representation}
Idea facets provide three primary advantages~\cite{fok2023scim, radensky2024scideator, pu2025ideasynth}: they are interpretable, reflecting how humans schematize paper content, they capture a paper's contributions and core structure, and they are useful as an abstraction to query LLMs, the output of which can be easily inspected and verified from the source text. 

We marshal facets into \textit{similarities} and \textit{deltas}, providing more granular units of transmission that signify an idea's development.
Similarities/deltas are also rooted in comparative analysis~\cite{pickvance2001four}, which we believe is central to understanding how papers relate to each other.
For example, Synergi~\cite{kang2023synergi} and Threddy~\cite{kang2022threddy} share \textit{similar} motivating problems (i.e., how to construct threads for scholarly synthesis), though Synergi provides an \textit{alternative} approach that shifts from manual thread construction to mixed-initiative curation, targeting the \textit{sub-problem} of manual overhead.
Extending this representation across a network of papers surfaces \textit{how} and \textit{why} new ideas evolve.
We summarize our motivations as follows:

\vspace{4px}\bpstart{M1: Provide useful bases of information.} Similarities represent shared intellectual ground, provisioning orientation and important context around a core set of ideas (Fig.~\ref{fig:teaser}A,G).

\vspace{4px}\bpstart{M2: Connect transitional forms of ideas.} Deltas can represent the morphing of a shared idea between papers (i.e., an idea ``basis'') into a new one. This transition can then propagate to other papers as a similarity, providing connective ``tissue'' of an idea from one paper to the next (Fig.~\ref{fig:teaser}B,F).

\vspace{4px}\bpstart{M3: Capture branching behavior.} Deltas materialize new branches / threads of ideas, enabling inspection of important developments or shifts in ideas across time (Fig.~\ref{fig:teaser}B,D,E).

\vspace{4px}\bpstart{M4: Attenuate irrelevant information.} Similarities implicitly represent a signal, indicating that an idea was important enough for future work to build on. Deltas may provide relevant context, either to make sense of what has been tried before or how it has been adopted in later papers.


\subsection{Pipeline}
Our pipeline consists of two key stages: deconstruction and collation.
References and citing papers over one hop of the citation graph are initially extracted for a seed paper.
The resulting set of papers are then used as input into the deconstruction phase.
We describe these stages and their outputs in detail in the following sections.
Prompts are provided as Supplementary Material.

\subsubsection{Deconstruction}
The deconstruction stage (Fig.~\ref{fig:pipeline}: Step 1) leverages our idea representation as input into an LLM system to extract salient similarities and deltas between \textit{pairs} of papers, parameterized by possible facets (e.g., problem, method, etc.).
We compare pairs of papers (Fig.~\ref{fig:pipeline}: Step 1C), using both papers' full text as contextual inputs to the LLM, to isolate fine-grained similarities/deltas, which can then be further aggregated into higher-level connections of ideas.
The LLM generates a set of key similarities and any deltas (Fig.~\ref{fig:pipeline}: Step 1D) and explicit links/transitions between them (Fig.~\ref{fig:teaser}E), which are then post-processed with a retrieval-augmented refinement loop (Fig.~\ref{fig:pipeline}: Steps 1E,F).

Specifically, extracted ideas from the initial step are first combined with the top-1 chunk (3-sentence window) of related text from the paper's vector store (using cosine similarity), along with surrounding paragraphs and the paper abstract/introduction as context.
We experimented with larger chunk sets and window sizes (e.g., top-12~\cite{fok2024qlarify} and top-4 chunks), though found that many of these chunks could be unrelated, degenerating to generic or incoherent outputs, or privilege idea capture for a particular paper.
Since our goal is to surface more conceptual relationships between ideas, this restriction helps ``force'' a specific concept to anchor on, using surrounding paragraphs for any additional nuance.
The LLM is then instructed to refine the idea similarity and/or delta based only on the provided chunks such that \textit{both} papers have verifiable attribution (\ie ideas are accurate and coherent, with useful grounding for reader inspection). 

\subsubsection{Collation}
The collation stage (Fig.~\ref{fig:pipeline}: Step 2) utilizes a sequence of clustering steps to construct an idea dependency graph, representing core units of analysis for synthesis and down-stream navigation in the system.
We partition ideas by citation direction (reference or citing paper) and the level of representation (similarity or delta), resulting in four separate layers (Fig.~\ref{fig:pipeline}: Step 2A).
Any papers not present in the original citation hull (e.g., other user-provided papers for tracing) are placed in the reference layer if published before the input paper or the citation layer if published after.
This partition enables users to both analyze the development of an idea (i.e., how citing papers build on ideas from the input paper and its references), as well as visualize shared ideas between related papers more visibly.

Each layer is clustered separately using agglomerative clustering, ensuring that all items are contained within a cluster for hierarchical navigation (Fig.~\ref{fig:pipeline}: Step 2B).
We apply a first height cut using the dynamic cuttree algorithm~\cite{10.1093/bioinformatics/btm563}, which balances the density of the resulting clusters to avoid highly uneven sizes.
We then cut the resulting clusters once more to generate sub-clusters, forming a 3-level hierarchy for improved organization (as in Synergi~\cite{kang2023synergi}).

The prior stage provides information about related ideas (either similarities or deltas) from a set of papers.
To trace these ideas \textit{between} layers, we perform an additional round of density-based clustering using HDBSCAN (Fig.~\ref{fig:pipeline}: Step 2C), which does not require heavy parameterization or that each item belong to a cluster.
Concretely, our goal is to first connect citing similarities (Fig.~\ref{fig:teaser}C) with any reference similarities (Fig.~\ref{fig:teaser}A,G), indicating core information bases shared across the network (\textbf{M1}).
We also care about connections between citing similarities and reference deltas (Fig.~\ref{fig:teaser}B,F), which represent transitional forms of ideas in the citing hull that build directly from an input paper's new delta/branch (\textbf{M2}).
Items from all layers are used as clustering input, except for citing deltas which already connect to the citing similarities.

We leverage clustering outputs to generate an idea dependency graph (Fig.~\ref{fig:pipeline}: Step 2D): nodes represent intra-layer clusters (Fig.~\ref{fig:pipeline}: Step 2B), and edges link all items from inter-layer clusters (Fig.~\ref{fig:pipeline}: Step 2C).
Edges are also populated between similarities and deltas from the same citing hull (Fig.~\ref{fig:teaser}E), which we leverage directly from deconstruction output (Fig.~\ref{fig:pipeline}: Step 1D) to capture branching behavior (\textbf{M3}).
Concise (3-8 word) labels and longer descriptions for each cluster and sub-cluster are generated with an LLM.
These labels are passed through an additional refinement stage to ensure minimal overlap and privilege more specific, grounded labels applying to \textit{all} items within a cluster.

\figurevlcasestudy

\subsection{User Interface}
We leverage the dependency graph generated by the LLM pipeline in an interactive system, visualizing idea traces as a node-link graphical layout.
Nodes represent clusters of shared ideas, with edges signifying transitions between two ideas from separate nodes (\eg similarity $\rightarrow$ delta).
Our design goals are as follows:

\vspace{4px}\bpstart{DG1: Surface related ideas from multiple papers.} 
A single idea may be used/scattered across multiple papers, often requiring citation chaining~\cite{palmer2009scholarly} or high-level overviews~\cite{taylor2022galactica, shahaf2012metro, heimerl2015citerivers, ma2025garden, gu2025abstractexplorer, fok2024qlarify} to locate and triage papers for further reading. 
Idea tracing should allow users to select/input ideas of interest and view their location across multiple papers at once.

\vspace{4px}\bpstart{DG2: Trace the development ideas across papers.} Critical to synthesis, idea tracing should allow users to determine how ideas mutate/accrete across papers to identify new branches or origins of interest.

\vspace{4px}\bpstart{DG3: Support hierarchical navigation across conceptual levels of ideas.} 
Certain users may be interested in high-level themes / topics, while others may be interested in variations on a particular idea (\ie how a specific method has been used or extended over time). 
Idea tracing should allow users to move through these different levels of abstraction.

\vspace{4px}\bpstart{DG4: Enable comparative analysis for papers outside the citation network.} 
Users may have papers/ideas they are familiar with, providing a useful starting point for synthesis, or wish to capture potential relationships between papers from different fields.
Idea tracing should allow users to ``map'' new papers onto the generated space, surfacing potentially novel connections to other papers within the input paper's citation graph.

\subsubsection{Selection and Tracing}
The interface provides a variety of selection mechanisms for inspecting clusters, papers, ideas within clusters, and facets of interest (\textbf{DG1,3}).
For example, selecting an upstream node from the reference similarity layer (Fig.~\ref{fig:case-study}A) reveals a detail view (Fig.~\ref{fig:case-study}B) with additional context from the papers that supports efficient verification (Fig.~\ref{fig:case-study}B).
Hovering or clicking an idea from the detail view propagates it across the dependency graph, enabling fine-grained inspection of an idea's influence/coverage over subsequent papers (\textbf{DG2}).

\subsubsection{Paper Injection}
To compare ideas from different communities (\textbf{DG4}), users can inject a paper by providing its unique identifier (\eg DOI) to the system (Fig.~\ref{fig:teaser}H).
The system fetches and parses the PDF, runs the deconstruction step against the input paper (Fig.~\ref{fig:pipeline}, 1C-F), and injects the results onto the extant dependency graph (Fig.~\ref{fig:injection}).
By default, the extracted ideas are placed in the reference or citing hull based on its publication date relative to the original input paper.
We do not re-cluster items, instead matching each newly deconstructed idea to the sub-cluster label with the highest cosine similarity (Fig.~\ref{fig:pipeline}: Step 1B).
To propagate across delta/similarity passthrough connections, we use the saved HDBSCAN clustering state (Fig.~\ref{fig:pipeline}: Step 1C) to predict cluster labels for the new items.

\figurenotebooklm

\section{Case Study: Tracing Vega-Lite}
We compare idea traces with Google NotebookLM~\cite{notebooklm}, a state-of-the-art LLM system for synthesizing and analyzing research ideas (Fig.~\ref{fig:case-study},~\ref{fig:notebook-lm}).
We select NotebookLM since it is representative of existing Q\&A workflows and allows us to upload the input paper's references/citing papers as a context store, providing for a more accurate comparison.
Below, we describe important affordances and insights with our approach, using the Vega-Lite~\cite{satyanarayan2016vega} paper as input.
We analyze this paper given its prominence in the visualization community, making it easier to evaluate output quality and demonstrate the value of idea traces.

\vspace{4px}\bpstart{Conceptual Drift of Ideas.}
By aggregating ideas across the paper network, idea traces can surface important shifts in topics over time, providing an overview of the paper's influence and subsequent developments.
For example, NotebookLM's output (Fig.~\ref{fig:notebook-lm}) generates \textit{finer} instances of key ideas from Vega-Lite (e.g., ``Selections''), enabled properties (e.g., ``Design Space Exploration''), and follow-up applications (e.g., ``Animation: Animated Vega-Lite'').
By comparison, idea traces visualizes the drift in ideas from low-level selection representations and multi-view coordination mechanisms (Fig.~\ref{fig:teaser}A) -- salient ideas that Vega-Lite draws on -- to coarser-grained applications (Fig.~\ref{fig:teaser}D) of specification repair, visualization recommendation, and new authoring tools and grammars.
This overview helps characterize the different roles that a paper may serve.
One explanation for Vega-Lite's influence is the resulting \textit{breadth} of new ideas and applications in the citing hull.

Idea traces also delineates \textit{causal} relationships more precisely. 
NotebookLM (Fig.~\ref{fig:notebook-lm}B) interleaves design exploration and recommendation systems at the same level of abstraction, while idea traces clarifies that recommendation (Fig.~\ref{fig:teaser}D) leverages Vega-Lite's structured representation for search and automation (Fig.~\ref{fig:teaser}(C)).

\vspace{4px}\bpstart{Fine-grained Lineages.}
In addition to providing poor granular representation for contextual overview, NotebookLM's output fails to characterize fine-grained relationships between ideas that are useful for understanding where a particular came from and branches of new ideas.
Take Vega-Lite's \textbf{Selection} abstraction as an example, which parameterizes a design space of \textit{interactive} graphics.
Idea traces connects this idea (Fig.~\ref{fig:case-study}C) to one of its precusors~\cite{wickham2010layered} that describes an expressive design space of \textit{non-interactive} graphics (Fig.~\ref{fig:case-study}A,B).
The branching/lineage of these two ideas are further delineated by ``passthrough'' connections (Fig.~\ref{fig:case-study}D,F).
In particular, work building on grammar-based design spaces (Fig.~\ref{fig:case-study}E) draws from this parentage thread (Fig.~\ref{fig:case-study}A), represented by the similarity passthrough (Fig.~\ref{fig:case-study}D).
Ideas that leverage Vega-Lite's structured specification (Fig.~\ref{fig:case-study}G) draw from both threads (Fig.~\ref{fig:case-study}A,C), indicating the prior influence of using grammars to specialize ``particular families of graphics'' (Fig.~\ref{fig:case-study}B) as well as Vega-Lite's concise representation that makes this enumeration tractable (noted by NotebookLM in Fig.~\ref{fig:notebook-lm}B).
Certain papers also leverage ideas specific to Vega-Lite's interaction design (Fig.~\ref{fig:case-study}H), indicated by one of the delta passthrough edges (Fig.~\ref{fig:case-study}F).

Finally, idea traces reveal clusters of new ideas from subsequent work (Fig.~\ref{fig:case-study}I-L), which can be traced back to their origins (Fig.~\ref{fig:case-study}A,C).
In contrast to NotebookLM's instance-level applications (Fig.~\ref{fig:notebook-lm}C), these clusters signify new ideas with similar lineages.
For example, the grammar of graphics' design space affordances (Fig.~\ref{fig:case-study}A) reveal specialized families of applications in line with the original motivation (Fig.~\ref{fig:case-study}B): linters and error-handling support (Fig.~\ref{fig:case-study}I); recommendation and animation systems (Fig.~\ref{fig:case-study}J); end-user authoring tools and scalable interaction (Fig.~\ref{fig:case-study}K); and new grammars that extend to other domains such as unit visualizations (Fig.~\ref{fig:case-study}L).

\vspace{4px}\bpstart{Influence Capture.}
Idea traces directly enable the discovery of influential precursors, which NotebookLM does not support.
Selecting different ideas at the top layer (Fig.~\ref{fig:teaser}A) immediately propagates their effect across the dependency graph, helping to evaluate influence.
For example, the root idea in Fig.~\ref{fig:case-study}A was chosen given its high coverage over subsequent ideas.
Idea traces also help characterize \textit{why} an idea was influential.
Design space expressivity (Fig.~\ref{fig:case-study}A,B), for instance, reveals 55 connections to citing papers (Fig.~\ref{fig:case-study}E) that directly note the enumerable properties of grammars.

Further, follow-up ideas from citing papers appear to relate to grammar extensions the most (22 connections in Fig.~\ref{fig:case-study}L) and interactive authoring tools the least (Fig.~\ref{fig:case-study}K).
One insight from this observation is how the visualization community may have responded to Vega-Lite's tradeoff of customization for concise specification, pivoting from authoring tools (which provided an effective alternative to the complexity of lower-level grammars~\cite{satyanarayan2014authoring, conlen2018idyll}) to new \textit{domain-specific grammars} that provide customization while retaining the benefits of conciseness.

\vspace{4px}\bpstart{Injection of Ideas.}
Pairwise analysis (Fig.~\ref{fig:pipeline}: Step 1C) enables the injection of any paper into a generated dependency graph (Fig.~\ref{fig:injection}).
This affords comparison with papers/ideas outside the immediate citation hull and can be used to surface key relations with progenitors. 
For example, NotebookLM's summary (Fig.~\ref{fig:notebook-lm}A) describes selections and view composition as core contributions of Vega-Lite; however, what precisely is the relationship between the two?
One option is to inject a paper that addresses both ideas, helping establish a clearer connection.
Exploiting Vega-Lite's overlap with the influential paper ``Brushing scatterplots''~\cite{becker1987brushing}, we can inject the latter since it is not directly cited.
The injected ideas reveal that Vega-Lite's \textit{selection} delta (Fig.~\ref{fig:injection}A) integrates two main prior threads, reframing brushing and linking (Fig.~\ref{fig:injection}B) as a selection mechanism (Fig.~\ref{fig:injection}C). 
This abstraction allows Vega-Lite to generalize SPLOMs (scatterplot matrices) to a compositional algebra (Fig.~\ref{fig:injection}A), resulting in a new family of interactive multi-view graphics.

\figureinjection

\section{Expert Evaluation}
To evaluate ideas generated by our pipeline and the utility of our system, our research questions were:
\begin{enumerate}
    \item[RQ1] How do researchers express key ideas for papers and how well do generated ideas capture them?
    \item[RQ2] What is the quality (accuracy, coherency, and relevancy) of generated ideas?
    \item[RQ3] What are the perceived benefits and drawbacks of using idea traces to assist with scholarly synthesis?
\end{enumerate}

\subsection{Procedure}
We recruited 10 researchers (all PhD students) to complete a 5-part evaluation study.
Each session lasted approximately 90 minutes and participants were compensated \$45.
Within each session, participants completed the following tasks:

\vspace{4px}\bpstart{Part 1 (10 minutes): Tutorial.} Participants were first provided an introduction to the tasks and idea traces system.

\vspace{4px}\bpstart{Part 2 (15 minutes): Idea Coverage.} 
Each participant provided a list of 3 papers that they had \textit{high} familiarity with (a paper from their research area that they had read thoroughly before) and contained at least 40 citing papers. 
The authors selected one of these papers before the study and cached the pipeline output.
Each participant was asked to first write down the key ideas for that paper, to avoid idea refinement across subsequent inputs, and then enter them into the system. 
Each idea was used to retrieve the top-10 ideas from cached output based on cosine similarity.
To assess alignment, participants were asked to select the \textit{minimum} number of ideas needed to sufficiently capture their input. 
To account for cases in which none of the selected ideas provided sufficient capture, participants then rated how well they believed the selected ideas captured their input idea using a 7-point Likert scale.

\vspace{4px}\bpstart{Part 3 (30 minutes): Idea Quality.} 
Participants analyzed the quality of a facet-stratified random sample of similarities and deltas from the same input paper as Part 2.
To capture nuance in idea quality, participants rated the accuracy, relevancy, and coherency of similarity and delta ideas using a 7-point Likert scale. 
In the results section, we provide different cutoff percentages and describe borderline instances.
When rating each similarity and delta, participants also examined extracted chunks used as supporting evidence (Fig.~\ref{fig:pipeline}: Step 1E) and had access to the full paper PDF for context.

\vspace{4px}\bpstart{Part 4 (20 minutes): Exploration.}
Participants completed an open-ended exploration task using our interactive system with Synergi~\cite{kang2023synergi}, a paper that none of them had read before. 
Participants were encouraged to explore different interactions/representations in the interface, including injecting papers that they had read before, to expedite synthesis, and tracing the development of ideas across the network. 
During this process, participants were also asked to think-aloud and describe things they liked, things that did not make sense, and vocalize how the information/navigation aligned with their typical workflow for literature review.

\vspace{4px}\bpstart{Part 5 (15 minutes): Interviews.}
We asked participants to reflect on their overall impressions and answer a set of survey questions on a 7-point Likert scale (see Appendix for full questions).

\subsection{Findings}

\subsubsection{RQ1: Capture of Input Ideas}
Participants enumerated 3-6 key ideas to test coverage ($\mu = 4.6, \sigma = 0.84$).
Some participants used keywords/phrases associated with high-level topics if they were discussed within a narrow context of the paper (e.g., P2 input ``prototyping'' for a paper on boundary objects~\cite{lee2005between}, which references this term as ``something that the observed people took part in, rather than, like, a design method'').
Others focused on specific facets of an overall idea (e.g., method or problem trying to be addressed) from their input paper.
For example, one of P3's inputs related to a specific \textit{mechanism} introduced in their paper, leading them to ignore any ideas not aligned on that dimension:

\begin{infobox}
    \textbf{Paper:} Improving Steering and Verification in AI-Assisted Data Analysis with Interactive Task Decomposition~\cite{kazemitabaar2024improving}
    
    \textbf{Input Idea:} Direct editing of assumptions
    
    \textbf{Selected Idea (Delta):} Makes assumptions directly editable, increasing user criticality and control instead of leaving them fixed or implicit.
    
    \textbf{Likert Score:} Sufficient Capture - 7 (Strongly Agree)
    
    \textbf{Participant Rationale:} ``I cared specifically about the interaction mechanism of editing assumptions, and this first one captures that mechanism sufficiently. Others bring up other ideas around it (like the challenges with propagation) that are brought up in the initial paper but were outside the scope of the concept as I cared about it mechanistically.''
\end{infobox}

For a given input idea, participants selected 1-9 ideas generated by our pipeline ($\mu = 3.8, \sigma = 1.87$), suggesting high sensitivity to the level of detail present in the input idea.
This nuance was captured by Likert scale ratings: participants felt that coverage was \textit{somewhat} sufficient (\text{$\geq$5} rating) for 74\% of their inputs, though only 52\% of instances had sufficient coverage (\text{$\geq$6}).
As we discuss in later sections, interactive loops that refine or disambiguate user queries may help alleviate this problem.

\subsubsection{RQ2: Quality of Generated Ideas}
Participants rated the quality of 110 ideas in total, split evenly between similarities and deltas.
Scores for partial alignment on similarity quality (\text{$\geq$5} rating) indicated that surfaced connections were useful, even when not fully precise: 82\% accuracy, 73\% coherency, and 76\% relevancy. 
However, high alignment (\text{$\geq$6}) was less prevalent: 67\% accuracy, 65\% coherency, and 60\% relevancy.
Two primary reasons for this differential included ambiguous language and muddled abstractions.
The following rating from P9 illustrates both cases:

\begin{infobox}
    \textbf{Input Paper:} Farsight: Fostering Responsible AI Awareness During AI Application Prototyping~\cite{wang2024farsight}
    
    \textbf{Reference Paper:} Improving Alignment of Dialogue Agents via Targeted Human Judgements~\cite{glaese2022improving}
    
    \textbf{Similarity:} LLM harms are addressed through downstream mitigation rather than relying only on pretraining-time interventions.
    
    \textbf{Input Paper Chunk:} ``These strategies identified by participants reflect some current strategies for mitigating LLM harms (§2.2).
    Looking ahead, we argue that it is crucial for future designers to provide users with harm mitigation suggestions and resources in systems similar to Farsight.
    Some participants in our study complained that Farsight is exploiting users' `empathy quota' and potentially desensitizing them about LLM harms, because Farsight only warns users about harms without providing mitigation suggestions (§6.6.3).''
    
    \textbf{Reference Paper Chunk:} ``In principle, this is a question that can be addressed empirically with suitable human experiments. 
    As discussed in section 7.3 of Rae et al. (2021), we believe most language harms are best mitigated downstream of LLM pretraining, due to faster iteration cycles, application-dependence of harms, and multiple roles served by a single model (we use Chinchilla as both policy and classifier). 
    This work is one component of this downstream mitigation, but our methods are limited to instance harms detectable by raters without significant help.''
    
    \textbf{Likert Scores:} Accuracy - 5 (Somewhat Agree); Coherency - 1 (Strongly Disagree); Relevancy - 6 (Agree)
    
    \textbf{Participant Rationale:} ``I think that the similarity statement [...] should show the opposite -- are addressed through pretraining-time intervention vice versa.''
\end{infobox}

\noindent First, the use of the word ``addressed'' in the similarity leaves room for interpretation: does this imply that both papers contribute \textit{methods} to address downstream mitigation (which is inaccurate, with both papers doing the opposite as noted in the participant's rationale), or that both papers address the problem in the literal sense, by discussing design \textit{strategies} for future systems that move beyond model training?
The extracted similarity further muddles the idea's abstraction by referencing model pretraining, which is only explicitly mentioned in the reference paper.
A clearer similarity might read, ``LLM harm mitigation is needed at the system/use stage beyond model training.''
Regardless, participants found that in many cases such as this one, useful information was still surfaced that could be re-interpreted from the provided chunks, as noted by P2 (``I thought it was very cool, surfaced things I would not have found on my own'') and P6 (``It uncovers the hidden/implicit connections between papers that sometimes even the author doesn't realize'').

Partial alignment scores for deltas (\text{$\geq$5} rating) were comparable to similarities: 80\% accuracy, 85\% coherency, and 76\% relevancy. 
The difference in coherency was notable: 85\% for deltas vs. 73\% for similarities.
As illustrated from the previous example, the primary reasons for this gap included vague abstractions and imprecise language needed to relate a similarity between two papers rather than one (in the case of deltas).
High alignment (\text{$\geq$6}) was also comparable: 71\% accuracy, 65\% coherency, and 71\% relevancy.
In this case, the difference in relevancy was notable: 71\% for deltas vs. 60\% for similarities.
We noted a tension between coherency and relevancy in several instances, often resulting in unbalanced scores.
Though a certain level of abstraction is needed to relate similarities between papers, producing higher coherence, this can come at the expense of reduced relevance to the main ideas of either paper, as noted by P2 (``Some of these statements, they feel like very general statements.'') and P3 (``I think one thought is this level of detail is extremely hard to [...] or it's, like, it's all quite vague.'').

\subsubsection{RQ3: Perceived Benefits \& Limitations of Idea Traces}
Satisfaction with idea traces was favorable ($\mu = 5.6, \sigma = 1.58$).
Participants reported finding relevant information ($\mu = 6.1, \sigma = 1.1$) that was often accurate ($\mu = 5.9, \sigma = 1.2$) and reliable ($\mu = 5.6, \sigma = 1.35$) from the surfaced paper chunks.
Participants provided a number of useful comments and suggestions, which we describe next.

\vspace{4px}\bpstart{Representation of Ideas.}
Several participants remarked on how both the layered organization and representation of ideas as similarities and deltas helped them think about papers more conceptually:
\begin{infobox}
    \textbf{P4:} ``I think I really like the very fact that you like chose an idea plane, and the papers are secondary, the ideas are evolving and mutating and changing, and papers are just representative.''
    
    \textbf{P6:} ``I think it's very intuitive, I immediately get it [...] because similarity is about what, what's the core of my paper or what's the core idea of a paper, and then delta is, how, like, what's critically new about this paper.''
    
    \textbf{P7:} ``Similarities and deltas, that made it easy to think about. Because I know exactly what I'm looking at and where it's coming from, whether it's past work or future work.''
\end{infobox}

\noindent Reflecting our initial motivations (\textbf{M4}), P6 noted the implicit signals afforded by the abstraction of similarities and deltas: ``I like how the similarity and delta, they attribute meaning to the connections [...] they're not just, like, connected, because they share the same keyword.''
Participants also appreciated that the tool motivated them to explore more ideas and connections ($\mu = 6.1, \sigma = 0.88$).
One participant's comment echoed prior systems that found additional benefits of improved curiosity and self-actualization~\cite{kang2023synergi}:
\begin{infobox}
    \textbf{P9:} ``But [ChatGPT] basically takes the fun out of doing related work [...] you basically, you don't really, like learn much. You don't know how an idea carries over from, like, a paper in 2000 to now, right?'' 
\end{infobox}

Participants also appreciated the hierarchical overview provided by the system, though some found certain cluster labels to lack specificity (P1, P7).
We observed a common usage pattern of starting at outer layers/nodes, clicking specific nodes to observe ``bottlenecks'' (as worded by P3), and then working inwards to specific ideas or papers within a node.
One participant (P4) leveraged the new ideas from the citing layer to work ``backwards'', starting with later nodes on the graph and reflecting back up to parent ideas to identify influential bottlenecks (an inversion of Fig.~\ref{fig:case-study}A).

\vspace{4px}\bpstart{Workflow \& Information Refinements.}
Participants found the tool's ease-of-use less favorable ($\mu = 4.2, \sigma = 1.87$), in large part due to the amount of information and visual complexity (P2-10).
Rather than entirely replacing the visual overview, which may be desired for complex sensemaking tasks (P2, P6, P7), participants remarked that a ``lightweight version'' (P6) would be more appropriate during the early stages of literature review.
Some participants recommended progressive disclosure features (P6, P9), mirroring systems like Qlarify~\cite{fok2024qlarify}, or the ability to filter out lower-level ideas (P4) or those with overwhelming jargon (P8).
Others desired a selection-based workflow like Synergi~\cite{kang2023synergi}: 
\begin{infobox}
    \textbf{P7:} ``Like, you have like a sentence, and you're like, oh, this is a new idea presented in the paper, I click on that sentence, and then it shows me, like, okay this is how prior work has inspired this idea, this is how future work has taken this idea and built on it.''
\end{infobox}
\noindent This workflow would align with desired use cases of tracking niche ideas (P1, P9) or assessing the novelty of a potential idea (P6).

However, one advantage of the visual overview, as noted by P7, is that it conveys the ``domain of the search space.''
Search interfaces like Google Scholar offer high recall (not domain-restricted), while Q\&A chatbots provide more precision at the expense of trust (i.e., important papers may be left out).
By integrating more lightweight features, idea traces could provide the benefits of both, allowing scholars to track provenance and visualize which papers influenced an idea, while also being able to restrict to more specific ideas or papers when desired.
Participants described several additional forms of metadata that would be useful to restrict the domain of search with idea traces: publication date/author lists (P9); representative ideas from figures (P6); and specific conferences (P1, P2, P7).

\vspace{4px}\bpstart{Idea-centric Applications.}
In line with our motivation for this work, participants were highly enthusiastic about the potential applications for idea traces.
They desired access to our tool when publicly released ($\mu = 6.3, \sigma = 0.82$), believing that it could add value to their research process ($\mu = 6.3, \sigma = 0.95$). 
We maintain that this is because \textit{ideas}, not \textit{papers}, are the core abstraction that scholars use throughout their activities, from reading and writing papers to teaching students and engaging colleagues.
However, the value of idea traces may extend from its ``orthogonality'' (P4) to standard methods for paper search and reading, with participants rating its alignment with the way they like to search and read papers lower than expected ($\mu = 4.7, \sigma = 2.26$).

One fruitful application participants noted would be to measure the influence and history of an idea more precisely, reflecting our earlier case study analysis (Fig.~\ref{fig:case-study}):
\begin{infobox}
    \textbf{P4:} ``So here is something that is not traditional, but I have seen this happen often: people, like, researchers from the same region, physical region, tend to cite their own peers, papers of their own peers, and the idea evolves in a very like, you know, local way, rather than being shared across the globe.''
    
    \textbf{P6:} ``[Is this idea] a follow-up to their previous idea, or has the idea been in the community for a while?''
\end{infobox}

\noindent Though our pipeline currently leverages the citation neighborhood for retrieval, it is citation-agnostic and can be executed with any set user-provided papers, using the reference/citing hulls as temporal partitions.
As referenced by P4, extending to papers outside the citation network could reveal interesting network-level insights obscured by citation filter bubbles~\cite{nguyen2014exploring, portenoy2022bursting}, such as structural holes~\cite{burt2018structural}.
In a similar vein, networks of ideas could be used as semantic operators (P3) to intersect ideas from different research communities or identify which ideas have \textit{not} been explored before.

Idea traces could also support annotations (P2) and personalization features such as generating ideas conditioned on what types of papers/ideas the user has encountered before (P9), analogous to personalized citations~\cite{chang2023citesee}.
Personalization could serve as a memory aid for revisiting old ideas/papers (P9), or function as boundary objects for collaborating with researchers across other disciplines (P2, P9).
Another useful application might be during the  writing or reviewing process (P6), ``auditing'' how well a new paper situates its contributions relative to prior work.

\section{Limitations \& Future Work}
Our evaluation demonstrates favorable capture and quality of generated ideas, as well as several downstream applications that could benefit from lifting the ``plane'' of synthesis from papers to ideas.
Deploying our system in real-world settings with both novice and expert researchers from a broader set of research fields can help better assess ecological validity.
Future work should also explore interactive refinement protocols (e.g., decomposing user input ideas for disambiguation and labeling) to provide better coverage of ideas.
For instance, users could engage with an initial concept alignment loop to personalize the underlying schema of similarities/deltas (e.g., abstract concepts represented by natural language descriptors~\cite{lam2024concept}, in lieu of concrete facets).
Finally, our pipeline requires PDF access to extract the full text for sufficient context, preventing traces to a vast amount of scientific papers behind paywalls.

\section{Conclusion}
We presented idea traces, a novel thread-based representation and system for extracting and collating ideas to trace their development across the literature.
The key \textit{delta} of this work lies in our representation of ideas as similarities and deltas, capturing salient relations between papers that provide useful context for scholars to interpret and synthesize new knowledge.
We evaluated our approach with experts through a set of tasks designed to gauge idea capture, quality, and usefulness, finding that generated ideas were high quality and generally provided good coverage.
Our findings surfaced several promising directions for future idea-based workflows, including mixed-initiative grounding loops and new forms of collaboration and personalization that move beyond the citation layer. 

\bibliographystyle{ACM-Reference-Format}
\bibliography{references}

\appendix

\section{Implementation Details}
Our pipeline leverages GROBID~\cite{grobid} for high-quality extraction of PDFs and GPT-5.4 for LLM-generated outputs due to its state-of-the-art reasoning capabilities. 
The interface is implemented with React (using react-flow for node-link layout) and uses a Flask server for deconstruction and collation at runtime.
Pipeline components are multi-threaded across papers to provide adequate processing times for a single input paper (approximately 5 minutes).

\section{Evaluation Details}

\subsection{Participant Demographics}
We provide participants demographics and background in Table~\ref{table:participants}.
\tableparticipants

\subsection{Idea Capture \& Quality Ratings}
We provide distribution plots for Likert scale ratings in tasks 2 (idea capture) and 3 (idea quality): Fig.~\ref{fig:capture-distribution},~\ref{fig:similarity-distribution},~\ref{fig:delta-distribution}.

\subsubsection{Rating Prompts}~\\
\bpstart{Similarity}:
\begin{itemize}
    \item ``Accurate (correctly describes a shared idea between both papers, without hallucinated or extraneous information)'': 1 - Highly Inaccurate, 7 - Highly Accurate. 

    \item ``Coherent (substantiated by clear reasoning and evidence from both papers, without overly vague language or unsupported interpretation)'': 1 - Highly Incoherent, 7 - Highly Coherent
    
    \item ``Relevant (meaningful, significant, or useful in the context of both papers)'': 1 - Highly Irrelevant, 7 - Highly Relevant.
\end{itemize}

\vspace{4px}\bpstart{Delta}:
\begin{itemize}
    \item ``Accurate (correctly describes a new idea in the later paper, without hallucinated or extraneous information)'': 1 - Highly Inaccurate, 7 - Highly Accurate. 

    \item ``Coherent (substantiated by clear reasoning and evidence from the later paper, without overly vague language or unsupported interpretation)'': 1 - Highly Incoherent, 7 - Highly Coherent
    
    \item ``Relevant (meaningful, significant, or useful in the context of the later paper)'': 1 - Highly Irrelevant, 7 - Highly Relevant.
\end{itemize}

\figurecapturedistribution
\figuresimilaritydistribution
\figuredeltadistribution
\figuresurveydistribution

\subsection{Interview Questions}
We asked participants the following questions during the last 15 minutes of each session:
\begin{itemize}
    \item Summarize your overall impressions of the idea traces interface and workflow. What did you like or dislike?

    \item ``How do you feel about the choice of representing ideas at more fine-grained levels (e.g., problems, methodologies, findings, etc.)? Are there other facets, dimensions, or representations you would have wanted?''

    \item ''How do you feel about the choice of connecting ideas between papers using `similarities' and `deltas'? Are there other representations you would have wanted?''

    \item ``How did you feel about the quality of extracted ideas (both similarities and deltas)?''

    \item ``How did you feel about the quality of traces/edges between connected ideas?''

    \item ``Do you feel that this workflow helped, worsened, or changed your approach to reading a new paper? If so, how?''

    \item ``What interactive features did you use? Did you find them useful? If so, how?''

    \item ``What other interactions, features, or improvements would you want in a future system for exploring ideas across scientific papers?''

    \item ``Do you think this approach would be useful in your own research workflows, such as literature review/search/exploration, and if so, how?''

    \item ``How does this system compare to existing tools/workflows (e.g., ChatGPT, Google Scholar, etc.) that you use often? What are the advantages/disadvantages of the idea tracing workflow compared to these existing methods?''
\end{itemize}

\subsection{Post-interview Survey}
Participants completed a usability survey after concluding their interviews (Fig.~\ref{fig:survey-distribution}).
All survey questions used a 7-point Likert scale (1 - Strongly Disagree, 7 - Strongly Agree):
\begin{itemize}
    \item \textbf{Satisfaction:} ``I felt satisfied with my experience using the tool to explore the ideas/papers.''

    \item \textbf{Relevancy:} ``The tool helped me gain relevant information while reading the ideas/papers.''

    \item \textbf{Accuracy:} ``The information gained through my exploration was accurate.''

    \item \textbf{Reliability:} ``The information gained through my exploration was reliable.''

    \item \textbf{Confidence:} ``I felt confident while exploring the ideas/papers.''

    \item \textbf{Motivation:} ``I felt motivated to ask questions or explore deeper into ideas/papers.''

    \item \textbf{Value:} ``I believe the tool can add value to my research process.''

    \item \textbf{Public Use:} ``If publicly available in the future, I would use a similar tool for exploring ideas/papers.''

    \item \textbf{Ease-of-Use:} ``I think learning to use the tool is easy.''

    \item \textbf{Research Fit:} ``The tool fits well with the way I like to search and read ideas/papers.''
\end{itemize}

\end{document}